\documentclass[12pt]{iopart}
\usepackage{iopams}  
\usepackage{setstack}

\newcommand{\bra}{\langle}
\newcommand{\ket}{\rangle}

\newcommand{\bv}[1]{{\boldsymbol #1}}

\begin{document}

\comment[Physics of Large Deviation]
{Physics of Large Deviation}

\author{Shin-ichi Sasa}

\address{Department of Basic Science, The University of Tokyo,
Tokyo 153-802, Japan}
\ead{sasa@jiro.c.u-tokyo.ac.jp}
\begin{abstract}
A large deviation function mathematically 
characterizes the statistical property of 
atypical events. Recently, in non-equilibrium 
statistical mechanics, large deviation functions 
have been used to describe universal laws such as the 
fluctuation theorem. Despite such significance, 
large deviation functions have  not been easily obtained 
in laboratory experiments. Thus, in order to 
understand the physical significance of large deviation
functions, it is necessary to consider their
experimental measurability in greater detail.
This aspect of large deviation is discussed 
with the presentation of a future problem. 
\end{abstract}


\section{Introduction}



hen observing a time series of some fluctuating quantity 
at discrete times, one may find an atypical event that occurs 
during a time interval. In many cases, atypical events might 
be negligible because of the low frequency  of their 
occurrence. However, there might be cases where 
such atypical events lead to a substantial effect,
as in the cases of seismic events, economic activities, 
and evolution of species. In such cases, it is significant 
to determine the frequency of atypical events. 


Atypical events are precisely formulated as follows.
For mathematical simplicity, a time series is assumed 
to be described by a discrete-time Markov chain
for a finite set of states.  The transition probability 
from $y$ to $x$ is denoted by $T_{xy}$.  The matrix $T$ 
satisfies $\sum_{x} T_{xy}=1$, and it is assumed to be 
irreducible. 
The initial state at time 0 is fixed  as $x_0$.
The probability of a trajectory 
$\omega=(x_1,x_2,\cdots, x_\tau)$ is expressed  as
\begin{equation}
{\cal P}(\omega)
=T_{x_\tau x_{\tau-1}}\cdots T_{x_1x_0}.
\end{equation}
Let $J_{xy}$ be a quantity defined at 
the transition to $x$ from $y$. The time average of
$J_{xy}$ during the time interval $[0,\tau]$ 
is expressed  as 
\begin{equation}
{\cal J}(\omega) = \frac{1}{\tau}\sum_{n=1}^\tau J_{x_nx_{n-1}}.
\end{equation}
Here, the law of large numbers is assumed. That is,
${\cal J}(\omega)$ almost surely converges to a value 
$J_*$ in the limit $\tau  \to \infty$. When ${\cal J}(\omega)$ 
deviates from the typical value  $J_*$, the time series
$\omega$ is considered to be atypical.
The probability of the deviation is low
for finite but large $\tau$; this probability is 
expressed as an exponential function of $\tau$:
\begin{equation}
{\rm Prob}(J) \simeq \exp(-\tau I(J))
\label{LD}
\end{equation}
for large $\tau$ \cite{Dembo,Ellis,Oono,Touchette}. 
This is called a large deviation property with 
a large deviation function $I(J)$.
The frequency of atypical 
events is characterized by $I(J)$.


At this point, let us recall that 
the fluctuation theorem \cite{FT}, 
which is a landmark in the recent development
of non-equilibrium statistical mechanics, states the
symmetry property of the large deviation function 
of the entropy production.
The fluctuation theorem holds for a  wide class of 
systems and  is  useful for deriving several non-equilibrium 
relations in a systematic manner. However, despite
universal validity, its applications are still 
limited to the formal aspects of non-equilibrium systems.


It should be noted that large deviation functions 
are hardly measurable in experiments, except in the
case of small fluctuations, because they characterize 
a small frequency of atypical events. For this reason,
physicists have not encountered large deviation functions 
in experiments.  
If the term ``physical quantity'' 
is used to describe a quantity  that can be 
measured in experiments, then  a large deviation function 
cannot be considered a physical quantity. 
Such disadvantage in the experimental measurability of 
large deviation functions may be related to limited 
understanding of the physical significance
of the fluctuation theorem. 
Thus, it would be a remarkable achievement 
if an experimental method for quickly obtaining $I(J)$ 
were to be devised. 
Recent development and future prospects of this achievement 
are briefly introduced in this Comment, as is implied by 
the title of this Comment, ``Physics of Large Deviation.''

\section{Basic idea}     %


The key idea of the experimental determination of $I(J)$
involves a biased ensemble expressed as
\begin{equation}
{\cal P}^h(\omega) \equiv 
\frac{e^{h\tau {\cal J}(\omega)} {\cal P}(\omega)}{Z(h,\tau)},
\label{bias}
\end{equation}
where the normalization constant $Z(h,\tau)$ is determined by
\begin{equation}
Z(h,\tau)=\sum_\omega e^{h\tau J(\omega)} {\cal P}(\omega).
\label{norm}
\end{equation}
The quantity $h$ in (\ref{bias}) is called 
a biasing field or a counting field.  
Here, the saddle-point calculation result for large $\tau$
is as follows:
\begin{eqnarray}
Z(h,\tau)& \simeq &  \sum_{J}e^{\tau(h J -I(J))}, \nonumber \\
& \simeq &  e^{\tau\sup_{J}(h J -I(J))}, \
\label{cg}
\end{eqnarray}
which leads to the definition of $G(h)$  as 
\begin{equation}
G(h)\equiv \sup_{J}[h J -I(J)].
\label{Gdef}
\end{equation}
This is the Legendre transform of $I(J)$. 
The inverse transformation is then expressed as
\begin{equation}
I(J)=\sup_{h}[h J -G(h)].
\label{invLeg}
\end{equation}
This expression indicates 
that the large deviation function $I(J)$ 
is equivalent to $G(h)$. 
$G(h)$ is called the scaled cumulant generating function,
because the $k$-th order derivative of $G(h)$ at $h=0$
is related to the $k$-th order cumulant of ${\cal J}$.


Now, suppose that the biased ensemble (\ref{bias}) 
can be generated experimentally.  
Let $\bra \ \ket^h$ be the expectation  with respect
to the biased ensemble (\ref{bias}). 
Then, $\bra {\cal J} \ket^h$ can be easily obtained in 
the experiments.  
Let $J_{\rm st}^h = \lim_{\tau \to \infty} \bra J \ket^h$.
From (\ref{cg}), the following equation is derived.
\begin{equation}
J_{\rm st}^h= \frac{dG(h)}{dh}.
\label{relate}
\end{equation}
The integration of this relation yields
\begin{equation}
G(h)=\int_0^h dh' J_{\rm st}^{h'}.
\label{integ}
\end{equation}
That is, $G(h)$ (and $I(J)$ through (\ref{invLeg}))
is determined by the measurement of $J_{\rm st}^{h'}$ for 
$0 \le h' \le h$. Thus, the  problem can be solved 
by generating the biased ensemble (\ref{bias}) in the experiments.


It should be noted that the biased ensemble can be generated 
in numerical experiments by a cloning method. (See Ref. \cite{Giardina} 
for a recent related study in non-equilibrium statistical mechanics.)
However, it cannot be performed in laboratory experiments.
Experimental operations are those of adding 
an additional force or changing the temperature,
which may be expressed by modifications of the transition matrix. 
Now, the problem becomes that of determining 
the modified transition matrix 
$T^h_{xy}$ that generates  the biased ensemble ${\cal P}^h(\omega)$
in the large $\tau$ limit. 

\section{Variational principle}     %


The transition matrix $T^h$ is characterized  by a  variational
principle. Its mathematical derivation is described below.
A set of transition matrices compatible with $T_{xy}$
is denoted by ${\cal V}_{T}$. (For $R_{xy} \in {\cal V}_T$, 
$R_{xy} >0$ only when $T_{xy} >0$.)
Let ${\cal R}(\omega)$ be the path probability generated by a
transition matrix $R$ in ${\cal V}_T$.
The application of Jensen's inequality to the trivial identity 
\begin{equation}
G(h)=
\lim_{\tau \to \infty} \frac{1}{\tau}\log
\sum_{\omega} 
{\cal R}(\omega)\frac{e^{h\tau {\cal J}(\omega)} {\cal P}(\omega)}
{{\cal R}(\omega)}
\end{equation}
leads to 
\begin{equation}
G(h) \ge
\lim_{\tau \to \infty} \frac{1}{\tau}
\sum_{\omega} {\cal R}(\omega)
\left [
h\tau {\cal J}(\omega)
- \log \frac{{\cal R}(\omega)}{{\cal P}(\omega)}
\right].
\end{equation}
Owing to the law of large numbers, this is rewritten as
\begin{equation}
G(h) \ge
\sum_{xy} R_{xy}p_y^{R}
\left[hJ_{xy} - \log  \frac{R_{xy}}{T_{xy}} 
\right],
\label{ineq}
\end{equation}
where $p^{R}$ is the stationary distribution
of the transition matrix $R$. (That is,
$\sum_y R_{xy} p_y^{R}=p_x^{R}$.)


On the other hand, (\ref{bias}) and (\ref{cg}) lead to
\begin{equation}
G(h)=
\lim_{\tau \to \infty} \frac{1}{\tau}
\left[h\tau {\cal J}(\omega)- 
\log \frac{{\cal P}^h(\omega)}{{\cal P}(\omega)}\right].
\label{bias2}
\end{equation}
Multiplying the both-hand sides by ${\cal P}^h(\omega)$ 
and taking the summation over all histories $\omega$,
one has 
\begin{equation}
G(h) =
\lim_{\tau \to \infty} \frac{1}{\tau}
\sum_{\omega} {\cal P}^h(\omega)
\left [
h\tau {\cal J}(\omega)- \log \frac{{\cal P}^h(\omega)}{{\cal P}(\omega)}
\right].
\label{g-15}
\end{equation}
Again, owing to the law of large numbers, 
(\ref{g-15}) is rewritten as 
\begin{equation}
G(h) =
\sum_{xy} T^h_{xy}p_y^{T^h}
\left[
hJ_{xy} - \log  \frac{T^h_{xy}}{T_{xy}}
\right] .
\label{eq}
\end{equation}
From (\ref{ineq}) and (\ref{eq}), the following variational
formula is obtained:
\begin{equation}
G(h) = \max_{R \in {\cal V}_{T}} \Phi(h,R)
\label{formula-1}
\end{equation}
and 
\begin{equation}
T^h=\underset{R \in {\cal V}_{T}}{\rm Argmax} \Phi(h,R),
\label{formula-2}
\end{equation}
where the function $\Phi(h,R)$ has been defined as
\begin{equation}
\Phi(h,R) \equiv \sum_{xy} R_{xy}p_y^{R}
\left[
hJ_{xy} - \log  \frac{R_{xy}}{T_{xy}}
\right].
\label{Phidef}
\end{equation}
Formula (\ref{formula-1}) is a well-known formula in 
large deviation theory. (See, for example, page 81 in Ref. 
\cite{Dembo} or page 284 in Ref. \cite{Ellis}. 
See also Ref. \cite{Andieux}, which reports
the formula.)


Furthermore, from the Perron-Frobenius theory 
for irreducible matrices, it is found that 
there exist a positive vector $\phi_x^*$ and a positive 
constant $\lambda$ such that 
\begin{equation}
\sum_x  \phi_x^* e^{h J_{xy}}T_{xy}  =\lambda \phi_y^*,
\label{restrict}
\end{equation}
where (\ref{norm}) and (\ref{Gdef}) lead to 
$G(h)=\log \lambda $. Then, 
with the definition 
\begin{equation}
R_{xy}^{h,\phi}\equiv
\phi_x T_{xy} e^{h J_{xy}} \phi_y^{-1}\mu 
\end{equation}
for any positive vector $\phi$, 
where the constant $\mu$ is determined as the normalization
condition $\sum_x R_{xy}^{h,\phi}=1$,
it is confirmed that $T^{h}=R^{h,\phi^*}$.
(Substitute $T^{h}=R^{h,\phi^*}$ into (\ref{eq}).)
The result indicates that 
\begin{equation}
G(h) = \max_{\phi>0} \Phi(h,R^{h,\phi})
\label{formula-3}
\end{equation}
and 
\begin{equation}
\phi^*= \underset{\phi >0 }{\rm Argmax}\Phi(h,R^{h,\phi}).
\label{Thvar}
\end{equation}
That is, the set of  variational parameters ${\cal V}_T$ 
is reduced to a set of positive vectors. 


In this manner, the transition matrix that generates the 
biased ensemble is characterized by the variational
principle. The best transition matrix to optimize 
the function $\Phi(h,R)$ can be obtained  by considering 
modifications to  the system. 
The formulas for continuous-time Markov jump processes 
and Langevin systems  can be directly obtained by 
considering an appropriate limit for (\ref{formula-3}) 
with (\ref{Phidef}).
The resulting formulas are equivalent to those 
reported in Refs. \cite{NemotoSasa1, NemotoSasa2}.
It should be noted that the formulas are considered to
be an extension of those  derived for systems with the 
detailed balance property \cite{garrahan,Jack}.


Now, the main message of Refs. \cite{NemotoSasa1, NemotoSasa2}
is that {\it the variational parameter $\phi_x$ corresponds 
to a potential function for physical models} that are
described by continuous time Markov jump processes 
and Langevin systems. 
That is, for these cases, the optimization can 
be considered in laboratory experiments. 
This indicates  that the biased ensemble is 
generated experimentally so that the large deviation 
function can be obtained  quickly in the experiments. 
The experimental determination 
of the large deviation function for a single 
Brownian particle under a non-equilibrium condition
was demonstrated in Refs. \cite{NemotoSasa1, NemotoSasa2}.

\section{Future problem}     %


The result appears to be very promising. 
However, the complication in the result
can be observed  immediately. 
In  the case of a single particle, only the external potential
needs to be controlled in order to obtain 
a biased ensemble; however,  the degrees of freedom
of dynamical evolution rules increase  exponentially
with the particle number.
For example,  when a system consisting of two particles
is considered,  one needs to modify the  interaction 
potential between the particles in addition to the 
modification of the  one-body potential.
In this regard, 
(\ref{Thvar}) is formally correct; however, 
it may not be applicable to many-body systems.


A one-dimensional lattice gas model consisting of $N$ 
sites in contact with  particle reservoirs at  two ends
is studied so as to investigate the problem more concretely.
The state of the system  
is denoted by $\bv{\sigma}=(\sigma_j)_{j=1}^N$, 
where $\sigma_j \in \{0,1\}$,
and the transition rate in the Markov jump process 
(or the transition matrix in the Markov chain) 
is expressed  as a $2^N \times 2^N$ matrix. Now, by using (\ref{formula-3}),
the optimization problem is formulated in the set of 
$2^N$ dimensional positive vectors. Each positive 
vector $\phi$ corresponds to a potential $V$ as 
a function of $\bv{\sigma}$. The potential can be
formally expanded in the form
\begin{equation}
V(\bv{\sigma})= \sum_{i} V_i^{(1)} \sigma_i +\sum_{ij} V_{ij}^{(2)}
\sigma_i\sigma_j+\cdots,
\label{expand}
\end{equation}
where, in principle, $k$-body, $N$-range interactions 
$V^{(k)}$ are included. The space of the variational parameters
is too large to be controlled.


Let us assume that experimentalists can control 
only a one-body potential for a macroscopic system;
this assumption corresponds to the case where 
only the first term in the expanded  form (\ref{expand}) 
can be considered for a system with large $N$.
An optimistic consideration  that the description of 
macroscopic behaviors does not require the complete 
information of microscopic details of the system leads 
one to expect that the control of a one-body potential 
is sufficient  to solve the optimization problem. On the other 
hand, cautious researchers would not have such an ill-founded 
expectation.  In any case,  the question may be 
mathematically stated as follows:
{\it Can the optimization problem  with a $2^N$-dimensional 
vector as a variational parameter be effectively reduced 
to an optimization problem  with an $N$-dimensional vector 
as a variational parameter in the macroscopic limit?} 


Obviously, such a reduction is not applicable to 
all systems. Then, an interesting future problem is 
to obtain  a condition under which the reduction indeed 
occurs. This problem may be related to the manner in which 
microscopic descriptions are connected with macroscopic 
descriptions in non-equilibrium systems. 
Hopefully, studying this future problem will provide 
a new direction to  non-equilibrium statistical mechanics.

\ack

The author thanks T. Nemoto for the collaborative work
on ``Physics of Large Deviation.''  The present study 
was supported by KAKENHI Nos. 22340109 and 23654130.

\end{document}